\documentclass[conference]{IEEEtran}
\IEEEoverridecommandlockouts
\usepackage{cite}
\usepackage{amsmath,amssymb,amsfonts}
\usepackage{algorithmic}
\usepackage{graphicx}
\usepackage{textcomp}
\usepackage{xcolor}
\usepackage{hyperref}

\usepackage{caption}
\usepackage{subcaption}

\def\BibTeX{{\rm B\kern-.05em{\sc i\kern-.025em b}\kern-.08em
    T\kern-.1667em\lower.7ex\hbox{E}\kern-.125emX}}
\begin{document}


\title{Intrusion Resilience Systems for Modern Vehicles}

\author{\IEEEauthorblockN{1\textsuperscript{st} Ali Shoker}
\IEEEauthorblockA{\textit{RC3, KAUST} \\
ali.shoker@kaust.edu.sa}
\and
\IEEEauthorblockN{2\textsuperscript{nd} Vincent Rahli}
\IEEEauthorblockA{\textit{University of Birmingham} \\
v.rahli@bham.ac.uk}
\and
\IEEEauthorblockN{3\textsuperscript{rd} Jérémie Decouchant}
\IEEEauthorblockA{\textit{Delft University of Technology} \\
j.decouchant@tudelft.nl}
\and
\IEEEauthorblockN{4\textsuperscript{th} Paulo Esteves-Verissimo}
\IEEEauthorblockA{\textit{RC3, KAUST} \\
 paulo.verissimo@kaust.edu.sa}
}

\maketitle

\pagestyle{plain} 


\begin{abstract}
Current vehicular Intrusion Detection and Prevention Systems either incur high false-positive rates or do not capture zero-day vulnerabilities, leading to safety-critical risks. In addition, prevention is limited to few primitive options like dropping network packets or extreme options, e.g., ECU Bus-off state. To fill this gap, we introduce the concept of vehicular Intrusion Resilience Systems (IRS) that ensures the resilience of critical applications despite assumed faults or zero-day attacks, as long as threat assumptions are met. IRS enables running a vehicular application in a replicated way, i.e., as a \textit{Replicated State Machine}, over several \textit{ECU}s, and then requiring the replicated processes to reach a form of \textit{Byzantine} agreement before changing their local state. 
Our study rides the mutation of modern vehicular environments, which are closing the gap between simple and resource-constrained "real-time and embedded systems", and complex and powerful "information technology" ones. It shows that current vehicle (e.g., Zonal) architectures and networks are becoming plausible for such modular fault and intrusion tolerance solutions---deemed too heavy in the past. Our evaluation on a simulated Automotive Ethernet network running two state-of-the-art agreement protocols (Damysus and Hotstuff) shows that the achieved latency and throughout are feasible for many Automotive applications. 

\end{abstract}

\section{Introduction}
\label{sec:intro}

Three trends, Automation, Digitization, and Connectivity are disrupting the ways modern vehicles are designed and used. While these trends can bring notable features like safety, efficiency, and convenience, they could turn into a curse if security and resilience are left as afterthoughts. 
Unfortunately, reality shows that safety and security incidents are doubling annually during the past three years, causing up to half Trillion dollars by 2024 due to cyberattacks~\cite{upstream-sec-rep:2022}, and leading to millions of car recalls~\cite{sw-eat-car:2021}. Such trend, if not contradicted, jeopardizes the sought features and puts human safety at risk~\cite{auto-safety-Koopman:2017}. We need novel approaches to improve vehicles' resilience: ensuring that an acceptable service prevails, even in uncertain environment conditions, or in the presence of faults or attacks that might not have been predicted (a.k.a, 0-days).

This work is motivated by two main observations in the automotive industry. The first is that the automation and digitization trends increase the complexity of vehicles and the likelihood of software faults and vulnerabilities. Digitization suggests software-defined vehicle systems (compute nodes, networks, and software) as a main enabler to automation, 
supporting features like x-by-wire, Advanced Driver Assistance Systems (ADAS), and Telematics.
This involves a considerable number of distributed software components running on over a hundred embedded compute devices, \textit{Electronic Control Unit}s (ECU), which communicate via in-vehicle networks, e.g., CAN bus, Automotive Ethernet, FlexRay, etc.~\cite{auto-eth-nets:2021}.
This results in a complex system with an enormous number---estimated to exceed 100 Millions---of Software Lines of Code (SLoC) in mainstream vehicles~\cite{sw-eat-car:2021,auto-market-mckinsey:2019}. Experience shows that human errors are positively correlated with both system's complexity and code footprint, and this increases the likelihood of benign faults and intrusions. 

The second observation is that connecting the vehicle to the cyberspace is becoming a mainstream. Connectivity is established in several networking forms like Vehicle to Everything (V2X), Cellular, 5G, Bluetooth, WIFI, GPS, or even through hardware memory sticks or USB connectivity~\cite{connected-apps:2016}. This raises substantial security challenges as it enlarges the attack surface and entry points of the vehicle system, and thus makes it highly prone to intrusions induced by (the well experienced) attackers in the cyberspace, via exploiting the existing vulnerabilities~\cite{net-security-survey:2013,lima2016towards}.

The automotive community has been recently focusing on consolidating the network security layer, leaving the higher software layers insufficiently addressed. Of particular interest is the introduction of new network security controls and tools (e.g., Gateways, Firewalls), and hardening the security of existing networks, e.g., FlexRay, CAN XL, Automotive Ethernet (100BASE-T1, 1000BASE-T1, and 10BASE-T1s), etc.~\cite{auto-eth-nets:2021}. This is also supported by using endpoint tools like \textit{Intrusion Detection Systems} (IDS) and \textit{Intrusion Prevention Systems} (IPS)~\cite{auto-cybersec:2021}. Nevertheless, IDS systems of either ``school''---signature-based and anomaly-based IDS---have limitations in the context of in-car systems, respectively blindness to zero-day vulnerabilities, and being difficult to define a ``normal behavior''. Not to mention the problem of real-time reaction/mitigation, which haunts IPS, and makes these ad-hoc  response techniques currently very limited (e.g., detaching a vulnerable ECU from the network bus using the Bus-off state~\cite{auto-cybersec:2021,IPS:2019}). In addition, since the network PHY/MAC protocols and tools (IPS/IDS) are application-agnostic, they can neither detect the anomalies and intrusions occurring at the upper layers nor stop their propagation to other ECUs.

In this paper, we introduce the concept of \textit{Intrusion Resilience Systems} (IRS) for modern vehicles. 
IRS aims at contributing to a timely revolution in current in-vehicle computer and network architectures, by extending the security and safety properties of component-based architectures (e.g., AUTOSAR). We propose SW-implemented fault and intrusion tolerance, leveraging available sets of failure-independent ECUs, e.g., multi-vendor Zonal ECUs with different AutoSAR implementations.
The approach is inline with the increasing demand for automotive computing and network channel redundancy, i.e., \textit{ASIL Decomposition}, as part of the ISO 26262v.2 safety standard~\cite{ISO26262,ASILdecompositionTexas}.

IRS is the first system-level automotive component that allows running multiple and possibly diverse replicas of a state-full application process on different ECUs, forming a resilient deterministic \textit{Replicated State Machine}~\cite{smr:1993}. Replicas are required to agree on a common state through a variant of \textit{Byzantine Agreement}~\cite{pbft:1999} protocols (today widely used in \textit{Blockchain}) prior to changing their local state. As long as the process is deterministic, agreement is reached despite the existence of benign or intrusion faults in a minority of replicas. Distributed applications like door locks, window control, software Over-the-Air (OTA) update verification are few examples on feasible applications on top of IRS.

IRS gives a quantum leap from IDS/IPS functions. First, it can work at a higher level of abstraction, targeting application software level anomalies and intrusions. Second, it follows an error masking approach which virtually captures all faults, even unknown ones, unlike IDS systems. Third, contrary to IPS whose response often degrades or suspends some system components or functions~\cite{auto-cybersec:2021,IPS:2019}, IRS makes it possible to roughly maintain the application functionality and quality under failures or attack.


In this work, we present a preliminary IRS Zonal architecture, and we drive a logical reasoning for its feasibility, given the recent technological advancements in modern vehicles. To demonstrate the concept, we apply it to a multi-vendor AutoSAR-based Zonal system, thus leveraging the diversity thereof, to improve the independence of failures of ECUs---which is a requirement for Byzantine agreement.  

We conducted an empirical evaluation for two state-of-the-art Byzantine agreement protocols, namely \textit{Damysus} and \textit{Hotstuff}---introduced in the Distributed Systems area. Our results show that IRS is feasible for modern Automotive Ethernet, since the achieved latency is less than 100ms for thousands of simultaneous operations. We argue that if more lightweight and efficient protocols are built especially for automotive, it is even possible to support time-critical applications.

The rest of the paper is organized as follows. Section~\ref{sec:irs} presents the concept and the architecture of IRS. Section~\ref{sec:discuss} analysis the feasibility conceptually, while Section~\ref{sec:eval} shows the empirical feasibility. The paper concludes in Section~\ref{sec:conc}.


\section{Intrusion Resilience System}
\label{sec:irs}

\subsection{Systems and Threat Models}
Consider an in-vehicle system of $N$ \textit{nodes}. A \textit{node} is composed of an computing device, i.e., an ECU, a corresponding software stack, and a (critical) soft real-time vehicular application for simplicity. (This can be generalized to many applications.) A node can communicate with its counterparts through messaging via a vehicular \textit{network}, either through a direct link, a switch, or via a gateway. A sent \textit{message} is assumed to eventually reach its destination node despite network failures or attacks (e.g., after re-transmissions). A node has a unique identity in the system to verify message authenticity and integrity using lightweight cryptography primitives, like \textit{Elliptic Curve Cryptography} (ECC). A node, or the application therein, is assumed to be \textit{deterministic}. However, an application can fail by crashing or behave arbitrarily or maliciously when subject to an intrusion. We assume that at most a fraction $F$ of $N$ nodes can fail at a time, which implicitly assumes some independence of failures between nodes. This can be achieved by employing ECUs from diverse vendors, different libraries, software stack, and implementation, etc., which is not uncommon in the automotive setting. Finally, we assume the existence of a technique to detect \textit{Denial of Service} (DoS) jamming attack in multi-hop bus networks like CAN and 10BASE-T1s~\cite{auto-eth-nets:2021,auto-cybersec:2021}.

\subsection{Architecture and Concept}

\paragraph{Concept}
The IRS concept is based on the idea of intrusion error \textit{masking} rather than detection and prevention as in IPS/IDS.
By running multiple ($N$) replicas/versions of an application and comparing their outputs on different nodes (ECUs), it is possible to mask any error caused by accidental or malicious faults occurring on $F$ faulty nodes, by adopting the output state of an uninfected majority ($N-F$). This is possible through running a Byzantine agreement protocol across application replicas. In this approach, the state of a critical application can only be modified upon the agreement of at least $N-F$ counterparts. This exploits the current replicated vehicle functionalities, often used for coordinated actuation and notification, to improve intrusion resilience.

\begin{figure}[t]
\centering
\includegraphics[width=0.8\columnwidth]{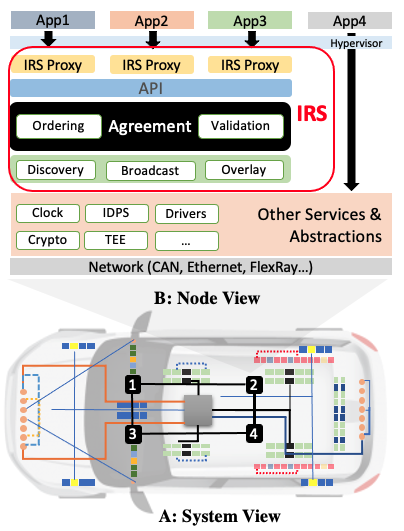}
\caption{Intrusion Resilience System (IRS) Architecture.}
\label{fig:IRS-arch}
\end{figure}

\paragraph{Architecture}
We present the IRS system view architecture in Fig.~\ref{fig:IRS-arch}, A. The System View shows a number $N$ of IRS nodes ($N=4$, in this case) replicated over $N$ ECUs. For clarity, we use \textit{Zonal Control Units (ZCU)} as ECUs to host different applications (e.g., door locks and window control) on the same ECU. 
On the other hand, Fig.~\ref{fig:IRS-arch}, B presents the Node View at one of the nodes (i.e., node 2) describing its components and relation within the Hardware/Software (HW/SW) stack.

In particular, the IRS is a system component, i.e., a module or service, used by those critical applications that require intrusion resilience. N versions of the application are employed over N different nodes, making use of the IRS module. The core module of the IRS seeks to ensure \textit{agreement} on requests issued by the application via an \textit{IRS proxy}. The proxy encapsulates the authentication, peer information, and the function to be made resilient through IRS in an application-agnostic way. The agreement module runs the main Byzantine agreement protocol to ensure (1) \textit{total ordering} on the application state and (2) output validation (i.e., comparison of results from counterpart nodes on other ECUs). The agreement module benefits from three underlying modules, namely, \textit{Discovery}, \textit{Broadcast}, and \textit{Overlay} to facilitate the membership management and networking with the peer nodes as a separate layer. Note that IRS can make use of these modules if made available by other frameworks, e.g., in the AutoSAR architecture.

IRS offers \emph{modular and incremental fault and intrusion tolerance} \cite{delta4:1988}. Not all node applications---or even functions of an application---are supposed to use the IRS, as they might not be critical, e.g., the case of App4 in the figure. Likewise, applications using IRS may resort to different models of replication (from crash to Byzantine fault tolerance), as well as different sizes of tolerance quorums (\#($N$)). For instance, an application that controls the remote door locks is much more critical than the mirror tilting application. Similarly, an Over-the-Air (OTA) update application is highly critical compared to infotainment social network (e.g., chatting) update.

IRS runs on top of other basic services and abstractions, such as those defined in the AutoSAR standard~\cite{autoSAR}. This way, it facilitates the integration of resilience in the existing component-based automotive architecture philosophy. 
At this layer, other tools like IDS, IPS may operate as well.
Finally, the bottom layer encapsulates the PHY network protocols (e.g., CAN, FlexRay, Automotive Ethernet) typically managed by the physical controller.
ECUs are connected via a network that could be multidrop, node-to-node, or switch-based network as long messages sent by one node are eventually delivered at the destination node.

\paragraph{Byzantine Agreement}
IRS encapsulates a distributed voting logic using an intrusion tolerant protocol category based on the concept of Byzantine Agreement/Consensus. 
Initial practical protocols~\cite{pbft:1999} would require $N=3F+1$, had quadratic ($O(n^2)$) messaging complexity, and were computationally demanding due to the heavy use of cryptography. The following generation was architecturally hybrid~\cite{hybrids-SGX}, featuring the use of trusted-trustworthy components~\cite{minBFT-2011efficient,correia2013bft}, dramatically reducing complexity, and requiring a smaller quorum of $N=2F+1$. Later, the advent of \textit{Blockchain} inspired yet another generation of intrusion tolerant protocols, becoming even more efficient and lightweight~\cite{damysus-BFT:2022}. The current state of affairs makes them feasible for environments with moderate capacities like modern vehicles (more on this in the next section). Describing a specific protocol is out of the scope of this position paper; however, we provide a brief overview of two recent protocols, namely Hotstuff and Damysus~\cite{yin2019hotstuff,damysus-BFT:2022}, that are used in the evaluation of IRS in Section~\ref{sec:eval}.

While the above protocols assume a \textit{partial-synchrony} network model, special real-time protocols are needed for bus networks like FlexRay and CAN. A good start is validating the two variants of Byzantine Resilient Real-Time protocols like \textit{PISTIS}~\cite{pistis:2021}. Unlike other intrusion tolerant protocols, which are non-synchronous, these real-time protocols are suited for hard or soft real-time environments as they have \textit{Timeliness} properties to guarantee delivery/execution given a defined probabilistic time-bound. 

\section{Feasibility Discussion}
\label{sec:discuss}
While the need for building resilient systems is very well understood, applying redundancy-based solutions like IRS may look infeasible for in-vehicular systems. 
Nevertheless, we argue that this is no longer the case as the three trends automation, digitization, and connectivity have changed modern vehicular systems dramatically. In this section, we try to alleviate these concerns by driving a conceptual analysis demonstrating the potential feasibility of IRS to modern vehicles.

\subsection{Distributed and Redundant Applications}
The current application landscape in automotive is very rich and complex, spanning ADAS \& Safety Systems, Infotainment, Body Electronics, Powertrain, and Telematics. At a fine-grained level, these applications incur millions of functionalities. For instance, a Volvo modern vehicle ``contains 10 million conditional statements as well as 3 million functions, which are invoked some 30 million places in the source code''~\cite{sw-eat-car:2021}. Many of these applications are becoming naturally distributed across the vehicle to manage the dependencies between functionalities and to synchronize the similar ones across the vehicle. For instance, a vehicle may have applications running four steering, braking, tyre pressure processes; four/five door lock and window processes; four light sets of processes, two mirror processes, several airbag processes, etc. Nevertheless, these processes are currently only synchronized in a passive  way, i.e., propagating notifications, where ``decisions'', e.g., changing an actuator state, are only made locally. Given this, the overhead of enforcing distributed control through agreement protocols prior to changing the application state would be reasonably low since replicas are already being used. This is sound for safety/security critical applications that are soft-real time, in particular, like door lock/unlock, window open/close, and OTA update validation by different processes on different ECUs. 

On the other hand, using redundancy to boost vehicle safety is becoming increasingly required~\cite{ASILdecompositionTexas}. Indeed, the \textit{ASIL Decomposition} mechanism drafted in the ISO 26262v.2 automotive safety standard~\cite{ISO26262,ASILdecompositionTexas} suggests using redundant computing nodes and network channels to improve safety and reduce the costs (e.g., by using redundant cheap nodes).

\subsection{Distributed Architecture}
The vehicular architecture has become heavily distributed as more ECUs are being added over time to cope with the application demands. Considering the evolution of distributed architectures~\cite{auto-market-mckinsey:2019}, applications are becoming more aggregated in larger ECUs: (1) Domain-based ones aggregate applications with similar functionalities; (2) Zonal-based ones aggregate based on the vehicle zone, e.g., a Door Control Unit hosts many applications (like door locks, motors, windows, theme lights) at the door proximity; and (3) Centralized. The former two are considered very convenient environments to run replicated protocols as the agreement protocol suggested in IRS. In addition, multiple aggregated applications can directly benefit from the IRS being a middleware/service. Indeed, while the replication cost has always been an adoption barrier in the IT world, the costs (surprisingly) look lower in vehicular architectures being natively distributed. 
The latter centralized architecture is getting more traction recently. We do not recommend this architecture from a security perspective, being a single point of failure/attack. Nevertheless, transforming the central controller into a distributed cluster could be a trade-off solution to mitigate this risk significantly.

\subsection{Efficient and Secure Networks}
Vehicular networks, especially the CAN bus, have always been considered slow and the weakest spot in a vehicle. In particular, the classical baud rate of CAN bus cannot be higher than 1Mbps, and the payload is only 8 bytes per packet~\cite{auto-eth-nets:2021}.
This prohibits an IRS-like solution where the agreement meta-data size (identifiers, signatures, cryptographic digests, clock) is high. 
On the other hand, CAN frames lack the sender/receiver identifiers which makes  authentication and integrity a non-trivial task. 
Nevertheless, as shown in the next table, the new versions of CAN, i.e., CAN FD and XL, have larger frame's payload size of 64B and 2KB, and baud rate to 2Mbps and 10Mbps, respectively. These are considered acceptable for soft real-time applications, e.g., like door locks and OTA updates, as response time is not critical. Furthermore, novel networks like Time-Triggered Ethernet (SAE AS6802), a.k.a., Automotive Ethernet and FlexRay have native security support and an order of magnitude higher bit rate. We believe that these advancements mitigate the concerns regarding the feasibility of IRS to such environment.

\begin{table}[h]
\label{tab:nets}
\caption{Modern automotive networking capabilities.}
\begin{tabular}{p{0.3\columnwidth}p{0.3\columnwidth}p{0.3\columnwidth}}
 \textbf{Network} & \textbf{MAX Baud rate} & \textbf{Max Frame size}\\\hline
 CAN-FD & 8Mbps & 64Bytes \\ 
 CAN-XL & 10Mbps & 2048Bytes \\ 
  FlexRay & 10Mbps & 254Bytes \\ 
 10BASE-T1 & 10Mbps & 1500Bytes \\ 
  100BASE-T1 & 100Mbps & 1500Bytes \\ 
   1000BASE-T1 & 1000Mbps & 1500Bytes \\ 
\end{tabular}
\end{table}


\subsection{Decent HW/SW Stack}
It can be assumed that running IRS agreement protocols in a constrained device (like a micro-controller-based ECU) and networks would be an overkill due to the heavy use of cryptography. Despite being challenging, modern automotive ECUs (microprocessor-based and multi-core) are getting high computational and storage capacities that could be compared to a \textit{Raspberry Pi} or a mobile phone\footnote{\url{https://www.emobility-engineering.com/focus-ecus/}}. This is correct, in particular, for main ECUs like domain and zone controllers, gateways, telecommunication units, etc. On top of this hardware, the software stack~\cite{e2e-market-mckinsey:2020} is also getting more mature while we observe more UNIX, POSIX, and Linux-based RTOS/OS, e.g., AGL, RTLinux, QNX, and Android Auto. This also means that a lot of IT/IoT libraries could now be adapted or used in automotive. New architectures are widely adopting the virtualization hypervisor technology, which facilitates application deployments on an ECU, and thus, replication in our case~\cite{e2e-market-mckinsey:2020}. Therefore, the modern HW/SW stack of modern vehicles is decent enough to support a solution like IRS.

\subsection{Diversity}
Independence of failures between replicas (i.e., ECU HW and SW) is a key challenge for the effectiveness of any Byzantine Agreement based system like IRS~\cite{castro2003base,garcia2014analysis,hypervisor-proactive,baudry2015multiple}. The reason is that without avoiding common-mode vulnerabilities or faults, many replicas can fail at the same time, thus violating the assumption of the correctness of a majority of replicas (N-f). While \textit{N-version} \textit{programming} is deemed an intuitive, but costly, approach to build software with independent implementations that have the same specification, it has been shown that diversifying the components, e.g., the operating systems or virtual machines, of the application's underlying layers is very effective to improve independence of failures~\cite{garcia2014analysis,hypervisor-proactive}. 

Leveraging this, we argue that diversity in automotive is less challenging than IT systems because of two reasons. First, the automotive SW/HW supply chain is big and multi-vendor, which provides a rich source of off-the-shelf black-box solutions to build diversity. For instance, it is not uncommon to have ECUs or MCUs of the same specifications, diverse software libraries, operating systems and hypervisors from many vendors. These are often used as underlying layers for applications to simplify their design and reduce the likelihood of leaving bugs or vulnerabilities. For a more conservative approach, the critical functions of an application can be chosen to run over IRS, which may optionally require only these functions to be implemented by different teams, e.g., using \textit{N-version} \textit{programming}. 
The second reasons is referred to the extensive use of standardized solutions in automotive. By design, this facilitates the integration of these modules as long as the APIs are defined and the specifications respected. We explain this by providing a case study using the AutoSAR standard~\cite{autoSAR}. \\

\begin{figure}
    \centering
    \includegraphics[width=0.8\columnwidth]{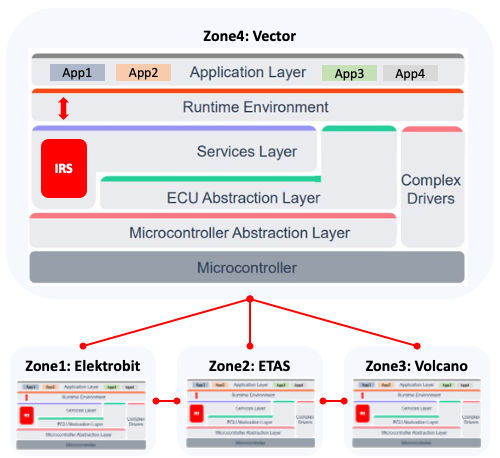}
    \caption{A zonal architecture of four Zonal Control Unit replicas with diverse AutoSAR implementations. The figure demonstrates how IRS can be integrated with AutoSAR to ensure intrusion resilience and leverage the AutoSAR standard to improve independence of failures between replicas.}
    \label{fig:autosar-diversity}
\end{figure}

\noindent \textbf{Case Study on AutoSAR.}
We provide a case study showing how to leverage the AutoSAR standard~\cite{autoSAR} to build diversity with intrusion resilience. In Fig.~\ref{fig:autosar-diversity}, we provide a possible integration of the AutoSAR architecture with IRS (depicted in a generic way in Fig.~\ref{fig:IRS-arch}). Particularly, we build a zonal architecture of four Zonal Control Unit (ZCU) replicas, Zone 1-4, that use different implementations of the AutoSAR specification at all layers: from the \textit{Microcontroller Abstraction Layer} at the bottom through the \textit{Runtime Environment} (a hypervisor) at the top. We select four different implementations (out of many~\cite{autoSAR-impl}) that are currently provided by well-known vendors following the AutoSAR standard. Each implementation is typically composed of up to more than 100 modules. This can generate a high level of diversity as it is less likely for one module to fail at the same time as its counterpart in the other ZCU replicas. Each Runtime Environment provides platform-agnostic access to the different modules and capabilities in the ECU Microcontroller Abstraction, ECU Abstraction and Services layers. The services layer is suggested as a good fit where the IRS modules are included. The same applications can run on top of the four ZCUs, whereas their agreement is ensured by the IRS modules and protocols. Notice that one can yet build more diversity by choosing different microcontrollers from different vendors at the HW layer as well.

\section{Evaluation}
\label{sec:eval}

The aim of this section is to drive an empirical evaluation to assess the feasibility of IRS to automotive networks and application requirements, i.e., throughout and latency. In particular, we evaluate the Byzantine agreement protocol, which is the most significant component of the IRS. Our goal is to show that the overhead of IRS is acceptable for some automotive applications even with current protocols, which are tailored for the IT heavyweight setting. 

\subsection{Brief summary on the protocols}

We consider two Byzantine agreement protocols as a baseline for our performance evaluation: HotStuff and Damysus. The former is chosen being a state-of-the-art fast protocol. The latter represents another class of protocols that can take advantage of hardware \textit{hybrids}, e.g., \textit{Hardware Secure Modules (HSM)}, common in modern ECU, to reduce the number of replicas needed and improve the performance. We concisely describe the main relevant features of these protocols, necessary to understand the evaluation. We refer the interested readers to learn about the protocols in~\cite{yin2019hotstuff,damysus-BFT:2022}. 

HotStuff~\cite{yin2019hotstuff} is a recent protocol optimized for high throughput. HotStuff's communication complexity is linear with the number of replicas/nodes, including special ones called \textit{leaders}. HotStuff requires $N \ge 3f+1$ nodes to tolerate $f$ Byzantine faults. Nodes build a chain of blocks (i.e., can be seen as batches) by voting for extensions, which are proposed by the leaders of \textit{views} (i.e., successive rounds).

Damysus~\cite{damysus-BFT:2022} is a hybrid BFT protocol that builds on HotStuff and leverages two trusted components, namely a \textit{checker} and an \textit{accumulator}. These can easily be implemented on modern trusted execution environments (TEE), because they only assume classical cryptographic functionalities and some memory. Therefore, these can be exploited in modern ECUs that often support HSM. The checker prevents nodes from equivocating, while the accumulator forces a leader to extend the most recent block. Thanks to these trusted components, Damysus uses only $N \ge 2f+1$ replicas, and requires one communication phase less than HotStuff. 
\subsection{Experimental Setting}
We evaluate here a version of basic HotStuff implemented in \textit{C++}. Replicas use \textit{ECDSA} signatures with \textit{prime256v1} elliptic curves (available in \textit{OpenSSL}), and are connected using the \textit{Salticidae} library. The protocol is deployed within Docker containers on a single machine equipped with an Intel Core i5-9500 CPU (3.00~GHz) with 6 cores and 32~GB of RAM. The network latency is enforced using \textit{netem}. The number of faults is set to be $1$ in all experiments, so a total of $4$ replicas, which are all directly connected with each other (i.e., no switched topology is used). The bandwidth varies between $10$, $100$, and $1000$~Mbps, simulating the bandwidth of 10BASE-T1, 100BASE-T1, and 1000BASE-T1 Automotive Ethernet networks~\cite{auto-eth-nets:2021}. In all experiments, we fix the network latency to $0.4$~ms, which is typical for Automotive Ethernet~\cite{zhu2021requirements}. We only consider Automotive Ethernet for three reasons: (1) it has high bandwidth that is suitable for heavy-weight agreement protocols; (2) it has similar synchrony model as the protocols we evaluate; and (3) it is believed that Ethernet will replace most in-vehicle networks in the near future. Our measurements focus on the latency and scalability of the protocols. The first measures the time for an ECU operation to complete, whereas the scalability shows the throughput limit of the protocol where latency remains acceptable under higher payloads.


\subsection{Latency}
In this experiment, we measure the latency of HotStuff's and Damysus's while varying the payload size, and blocks contain a single transaction with payload of size $8$, $128$, or $1024$~Bytes (B). In addition to the payload, a transaction contains $2\times{4}$~B for metadata (a client id, and a transaction id), as well as the hash value of the previous block of size $32$~B, thereby adding $40$~B to each transaction in addition to its payload. Therefore, given the above payloads, each
transaction is of size $48$, $168$, or $1064$~B. Each experiment presents the average of $10$ repetitions with $30$ views each (so a total of $300$ instances). 

\begin{figure}[t]
    \centering
    \includegraphics[width=.9\columnwidth]{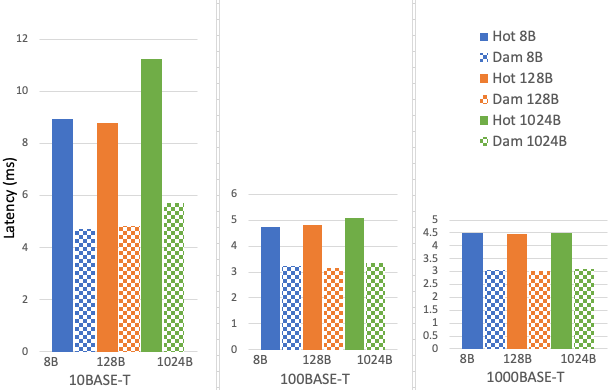}
    \caption{Latency ($ms$) of Hotstuff (Hot) and Damysus (Dam) with simulated Automotive Ethernet and varying payload size and link latency $400 \mu s $.}
    \label{fig:lat}
\end{figure}

Fig.~\ref{fig:lat} presents HotStuff's and Damysus's latencies depending on the bandwidth of the various Automotive Ethernet ($10$, $100$ or $1000$~Mbps) and depending on the payload size ($8$, $128$ or $1024$~B). In this scenario, we evaluate the protocols' latency under minimal workload to measure the lowest possible latency. Our measurements indicate that a request can always be treated in between $8$ and $12$~ms for Hotstuff and between $4$ and $6$~ms for Damysus, with a bandwidth of $10$~Mbps. Increasing the Ethernet bandwidth decreases the protocols' latency. Requests are respectively processed in less than $5.1$ and $4.51$~ms for Hotstuff, with a $100$~Mbps and $1000$~Mbps. Damysus's latency is as low as $3$~ms in both 100BASE-T1 and 1000BASE-T1. We expected Damysus to have lower latencies because the use of HSM abstractions reduces the message exchange round-trips. 

These results are considered very acceptable latency numbers for many Body, Chassis, and Power-terrain applications.  Larger requests should logically increase HotStuff's latency, which is the case for our experiments with $100$~Mbps; however, this effect is difficult to observe under a low workload and high bandwidths. 

\subsection{Scalability}

\begin{figure*} [t]
     \centering
     \begin{subfigure}[b]{0.31\textwidth}
    \centering
    \includegraphics[width=\columnwidth]{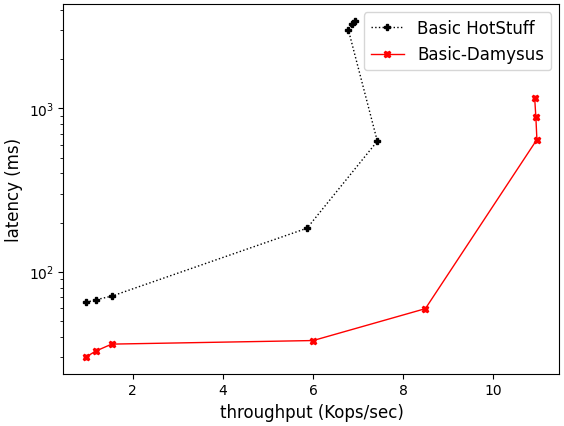}
    \caption{10BASE-T1}
    \label{fig:sca-10}
     \end{subfigure}
     \hfill
     \begin{subfigure}[b]{0.31\textwidth}
    \centering
    \includegraphics[width=\columnwidth]{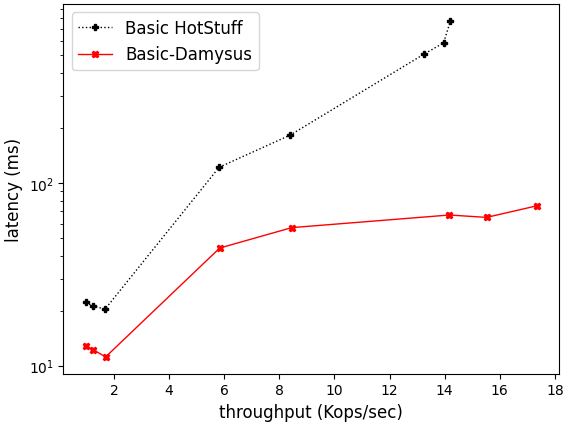}
    \caption{100BASE-T1}
    \label{fig:sca-100}
     \end{subfigure}
     \hfill
     \begin{subfigure}[b]{0.31\textwidth}
    \centering
    \includegraphics[width=\columnwidth]{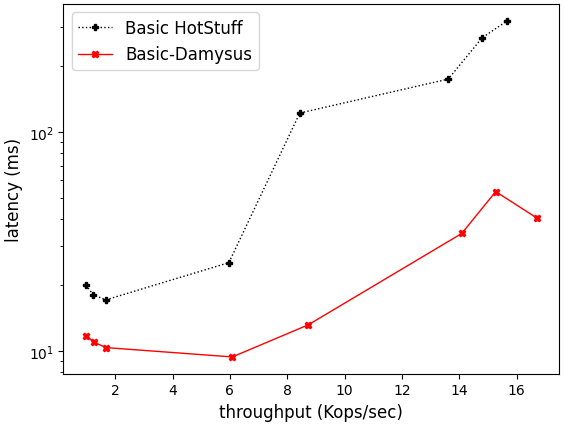}
    \caption{1000BASE-T1}
    \label{fig:sca-1000}
     \end{subfigure}
        \caption{Scalability with simulated Automotive Ethernet}
        \label{fig:three graphs}
\end{figure*}

We then study the influence of the system's workload on HotStuff and Damysus' throughput and latency in Fig.~\ref{fig:sca-10}, Fig.~\ref{fig:sca-100} and Fig.~\ref{fig:sca-1000}, corresponding to 10BASE-T1, 100BASE-T1, and 1000BASE-T1, respectively.
In these experiments, we increase the rate with which clients submit requests; hence the curve points in the figures correspond to the delays between issuing subsequent operations: $900,700,500,100,50,10,5,0$~microseconds. 
These rates can represent the load on the IRS, i.e., where multiple applications are running simultaneously. In addition, running two clients considers the cases of concurrent views---which could incur some race conditions. Blocks/batches are composed of $400$ transactions/operations, with $0$~B payloads (again, plus $40$~B for the above information).
In all experiments, the protocol's throughput and latency increase with the request rates until the system saturates. Under perfect settings, it is expected that the protocol latency increases exponentially and that the throughput remains constant.

HotStuff's maximum throughput with $10$, $100$ and $1,000$~Mbps networks is roughly $7$, $14$ and $16$~Kops/sec, respectively. More interestingly, it scales up to $4$, $6$ and $8$~Kops/sec while acheiving latency less than $100$ms. In general, under the same settings, Damysus' throughput is higher at around $11$, $17$ and $17$~Kops/sec; and $9$, $18$, and more (the network is not saturated here), while keeping a latency less than $100$ms. These are very promising results, showing that the network serves thousands simultaneous critical applications like Door locks, OTA firmware/software update, etc. 
Both throughput and latency improvements are expected because Damysus has one communication phase less than HotStuff. 

These results indicate that using an IRS for vehicles is possible with the recent advancements in vehicle networks and controller capabilities. This encourages more research on devising Byzantine agreement protocol variants that are more automotive-friendly. A promising directions seems taking advantage of the HSM hybrid to build more efficient and lightweight protocols for Automotive Ethernet. Other protocols may also be built for multi-hop networks like CAN-XL and FlexRay. This requires more work on the network synchronization modeling, that may benefit from real-time Byzantine broadcast protocols~\cite{pistis:2021} that ensure a notion of \textit{timelines}, useful for safety-critical applications.



\section{Conclusion}
\label{sec:conc}
We introduced the concept of\textit{ Intrusion Resilience Systems} (IRS) for modern vehicles. The aim is to bridge the gap left in security-by-design and intrusion detection and prevention systems at two levels: first, it is tailored for the software/application layer; second, it tolerates faults and 0-day attacks to roughly maintain the same service quality even if intrusions could not be profiled. IRS uses the \textit{State Machine Replication} approach in which the replicated application can only change the local state upon \textit{Byzantine agreement} with its counterpart nodes. The paper proposed a preliminary architecture and an analytic feasibility study that highlights the fact that modern vehicular technologies are closing the gap with IT/IoT technologies, which makes them plausible environments to adopt a replicated solution as IRS. 
The results of our empirical evaluation using two state-of-the-art protocols, \textit{Damysus} and \textit{Hotstuff}, 
shows that IRS is feasible for modern Automotive Ethernet,
since the achieved latency is less than 100ms for thousands
of simultaneous operations.
We invite researchers and practitioners to investigate this direction by studying the tradeoffs of agreement protocols, architectures, diversity, application space, etc.



\bibliography{ref.bib}
\bibliographystyle{plain}


\end{document}